\documentstyle[prd,eqsecnum,preprint,tighten,aps]{revtex}
\newcommand{\sump}{\sum_{n=0}^\infty \!{}^{'}}

\newcommand{\ve}{\varepsilon}
\begin{document}
\draft
\title{Casimir effect for a dilute dielectric
ball at finite temperature}
\author{V.V.~Nesterenko\thanks{Electronic address: nestr@thsun1.jinr.ru}}
\address{Bogoliubov Laboratory of Theoretical Physics,
Joint Institute for Nuclear Research, 141980 Dubna,  Russia}
\author{G.~Lambiase\thanks{Electronic address: lambiase@sa.infn.it}
and G. Scarpetta\thanks{Electronic address: scarpetta@sa.infn.it}}
\address{Dipartimento di Scienze Fisiche ``E.R. Caianiello'',
 Universit\'a di Salerno, 84081 \\ Baronissi (SA), Italy.\\
 INFN, Sezione di Napoli, 80126 Napoli, Italy.\\
and International Institute for Advanced Scientific Studies--Vietri sul Mare,
Italy.}
\date{\today}
\maketitle
\begin{abstract}
The Casimir effect at finite temperature is investigated for a dilute
dielectric ball; i.e., the relevant internal and free energies are
calculated.  The starting point in this study  is a rigorous general
expression for the internal energy of a system of noninteracting
oscillators in terms of the sum over the Matsubara frequencies.
Summation over the angular momentum values is accomplished in a
closed form by making use of the addition theorem for the relevant
Bessel functions. For removing the divergences the renormalization
procedure is applied that has been developed in the calculation of the
corresponding Casimir energy at zero temperature.  The behavior of
the thermodynamic characteristics in the low  and high temperature
limits is investigated.
\end{abstract}

\section{Introduction}
The calculation of the  vacuum  electromagnetic energy of a dielectric
ball has a rather long history~\cite{M,MNg1,BNP}. Only recently was
the final result obtained for a dilute dielectric ball at zero
temperature. Remarkably, this Casimir energy was calculated
by different methods: (i) by summing up the van der  Waals forces
between the individual molecules inside the compact
ball~\cite{MNg2}; (ii) in the framework of a special perturbation
theory, in which the dielectric ball was treated as a perturbation
when considering the electromagnetic field in unbounded empty
space~\cite{Barton}; (iii) by making use of the Green's functions
of the quantized Maxwell field with an explicit account of the so
called contact terms \cite{BM,BMM,Mar},
 on the stage of the numerical calculations
the uniform asymptotic expansion for the Bessel functions and the
zeta regularization technique being applied; (iv) by the mode
summation method with the use of the addition theorem for the
Bessel functions~\cite{LSN}. In calculations without using the
uniform asymptotic expansions for the Bessel functions
\cite{MNg2,Barton,LSN} the exact (in the $\Delta
n^2$ approximation) result for the Casimir energy under
consideration was obtained. The general structure  of the
ultraviolet divergencies in this problem was clarified  in Ref.\
\cite{Bordag}.

Undoubtedly, it is interesting to extend these theoretical studies to
finite temperature. It is worth noting here that the total number
of papers concerned with the calculation of the Casimir effect at
finite temperature, and especially for spherical boundaries, is not
so large. First Ref.\ \cite{BD} should be mentioned, where the
multiple scattering expansion has been developed in order to
investigate the vacuum effects for perfectly conducting boundaries.
The calculation of the vacuum electromagnetic energy of a compact
ball with the same velocity of light inside the ball and in the
surrounding medium proves to be not more involved.  In papers by
Brevik and co-authors \cite{BNP,BC,BC1,BC2,BC3,BY} this problem has
been studied at zero temperature and at  finite temperature, as well
as with allowance for dispersion (see also Refs.\
\cite{NPLett,LNB,Klich,KFMR}). However, the Casimir effect at finite
temperature for a dielectric ball made of nonmagnetic material has
not been considered till now.

An essential advantage of the calculation of the Casimir energy of a
dilute dielectric ball, carried out in Ref.\ \cite{LSN} by the mode
summation method, is  the possibility for its straightforward
generalization to the finite temperature. It is this problem that
will be considered in the present paper.  The employment of the
addition theorem for the Bessel functions enables one to carry out
the summation over the angular momentum in a closed form. As a
result, the exact (in the $\Delta n^2$ approximation) value for the
Casimir internal and free energies of a dilute dielectric ball will
be derived for finite temperature also. The divergencies,
inevitable in such studies, will be removed by making use of the
renormalization procedure developed previously for calculation of the
relevant Casimir energy at zero temperature. Both thermodynamic
characteristics are presented as the sum of the respective quantity
for a compact ball with uniform  velocity of light and an additional
term  which is specific only for a pure dielectric ball. The behavior
of the thermodynamic characteristics in the low  and high
temperature limits is investigated. The low temperature expansions
for the internal and free energy involve, except for the constant
term, only even powers of the temperature $T$ beginning from $T^4$.

The layout of the paper is as follows. In Sec.\ II the general
formulas are introduced for the internal and free energies of a
system of noninteracting oscillators in terms of the sum over the
Matsubara frequencies. This enables one to avoid the ambiguities
arising when the formal substitution of the integration over the
imaginary frequencies by the summation over the discrete Matsubara
frequencies is used in the integral representation for the
relevant Casimir energy at zero temperature.  The renormalization
procedure needed for removing the divergencies is also discussed
here. In Sec.\ III first the internal Casimir  energy
of a dilute
dielectric ball is calculated. Next the free energy is obtained by
partial integration of the relevant thermodynamic relation. The low
and high temperature limits of the thermodynamic characteristics are
examined. In the Conclusion (Sec.\ IV) the obtained results are
summarized and future studies in this area are outlined.
\section{Transition to the finite temperature in calculations of the Casimir
energy}
 Usually the transition to finite temperature in calculations
of vacuum energy is accomplished by substituting the integration over
imaginary frequencies by summation over the discrete Matsubara
frequencies in  the integral representation for the Casimir energy
at zero temperature \cite{BY,Brevik,LP}. However, following this
way one should control how many times the integration by parts in
the integral expression at hand has been done \cite{NP,Kpc}. The
point is that in this way one may obtain both the internal energy
and the free energy at finite temperature of the system under
consideration. The corresponding examples can be found in Ref.\
\cite{NP}.

Keeping this in mind, we shall proceed from the general formulas
determining the internal energy and the free energy of a set of
noninteracting oscillators at finite temperature. Here we briefly
remind the readers of
these formulas and their simple derivation.
  The natural system of units is used, where $c=\hbar=k_{\text B}=1$,
$k_{\text B}$ being the Boltzman constant.

Let us consider an infinite set of  noninteracting oscillators with
frequencies determined by the equations
\begin{equation}
\label{freq}
f_{\{p\}}(\omega,a)=0\,{.}
\end{equation}
Here $a$ denotes some parameters specifying the configuration of
the system at hand and $\{p\}$ is a complete set of quantum
numbers characterizing the spectrum. In the case under
consideration $a$ stands for the radius of the ball, and $\{p\}$
incorporates the orbital ($l$) and azimuthal ($m$) quantum numbers
and the type of the solutions of the relevant Maxwell equations
[the transverse electric (TE) and transverse magnetic (TM) modes].
The free energy of such a set of noninteracting oscillators at
finite temperature $T$ is determined by the formula
\begin{equation}
\label{free}
F(T)= T \sum_{\{p\}} \sump \ln f_{\{p\}}(i\omega_n,a)\,{,}
\end{equation}
where $\omega_n$ are the Matsubara frequencies
\begin{equation}
\label{Matsubara}
\omega_n=2\pi n T,
\end{equation}
and the prime on the summation sign means that the $n=0$ term is counted with
half weight.

Derivation of the formula (\ref{free}) can be found in Ref.\ \cite{NP}, where
the free energies of  individual oscillators
\begin{equation}
\label{free1}
F_1(T,\omega)=\frac{\omega}{2} +T\ln\left (
1-e^{-\omega/T}
\right)
\end{equation}
have been summed by the contour integration method. In Eq.\ (\ref{free1})
$\omega$ are the roots of the frequency equations (\ref{freq}). It is assumed
that for a given set $\{p\}$ Eq.\ (\ref{freq}) has an infinite countable
sequence of the roots. In order to find the sum of the free energies
(\ref{free1})
corresponding to this sequence of the roots, in paper \cite{NP}
the contour integration has been
used.

Having obtained the free energy (\ref{free}) one can derive the internal energy
$U(T)$ of the set of noninteracting oscillators  by making use
of the thermodynamic relation
\begin{equation}
\label{td}
U(T)=\frac{\partial }{\partial \beta}\left [
\beta F(T)
\right], \quad \beta = T^{-1}\, {.}
\end{equation}
Substitution of Eq.\ (\ref{free}) into this relation gives
\begin{equation}
\label{internal}
U(T)=-T\sum_{\{p\}}\sump\omega _n\frac{d}{d\omega _n} \ln
               f_{\{p\}}(i\omega_n,a)\,{.}
\end{equation}

Formula (\ref{internal}) can be derived directly by summing the internal
energies of the individual oscillators
\begin{equation}
\label{interna1}
U_1(T,\omega)=\frac{\omega}{2}\coth \left (\frac{\beta \omega}{2}
\right )
\end{equation}
and applying for this purpose the contour integration \cite{KFMR}.
In this case the sum  over the Matsubara frequencies
(\ref{Matsubara})  in Eq. (\ref{internal})  arises
as a result of evaluation of the respective contour integral by the residue
theorem, the residues   being taken at the poles of the function $\coth(\beta
\omega/2)$.

Certainly, the representations (\ref{free}) and (\ref{internal})
are formal because they involve the divergencies. Therefore in
order to obtain the physical results, an appropriate
renormalization should be done. This procedure includes
specifically the subtraction of the vacuum energy of unbounded
homogeneous space \cite{PRep,MT}. This can be achieved by the
following  substitution in Eqs. (\ref{free}) and
 (\ref{internal})
\begin{equation}
\label{subtr}
f_{\{p\}}(i\omega_n,a) \to \frac{f_{\{p\}}(i\omega_n,a)}{f_{\{p\}}(i\omega_n,
\infty )}\,{.}
\end{equation}

In Ref.\ \cite{LSN}  it was shown that in the case of a
dielectric ball an additional renormalization is to be done,
namely, the contribution into the vacuum energy, which is
proportional to $\Delta n$, should also be subtracted. Here
$\Delta n = n_1-n_2$, with $n_1\;(n_2)$ being the refractive index
inside (outside) the ball. We shall follow this scheme  at finite
temperature too (see the regarding reasoning in the next section).

\section{Internal and free energies of a dilute dielectric ball at finite
temperature}
We shall consider a solid ball  of radius $a$ placed in an unbounded uniform
medium, the temperature $T$ of the ball and of the ambient medium being the
same. The nonmagnetic materials making up the ball and its surroundings are
characterized by permittivity $\varepsilon_1$ and $\varepsilon_2$,
respectively. It is assumed that the conductivity in both the media is zero.
 Further we put
\begin{equation}\label{dn}
  \sqrt{\varepsilon_1}=n_1=1+\frac{\Delta n}{2}\,,\quad
 \sqrt{\varepsilon_2}=n_2=1-\frac{\Delta n}{2}
\end{equation}
and  assume that $\Delta n<<1$. From here it follows, in particular, that
\begin{equation}\label{epn}
  \varepsilon_1-\varepsilon_2=(n_1+n_2)(n_1-n_2)=2\,\Delta n\,.
\end{equation}

     In the problem at hand, as well as in  the
other ones (see the examples in Ref.\ \cite{NP}), it is convenient to
calculate first the internal energy of a dielectric ball using Eq.\
(\ref{internal}) and then to get the free energy by integrating the
thermodynamic relation (\ref{td}).

Equations, determining the frequencies of the electromagnetic oscillations
 associated with a dielectric ball, are \cite{Stratton}
\begin{equation}\label{dd}
  \Delta_l^{\text{TE}}(a\omega)=0\,,\quad
  \Delta_l^{\text{TM}}(a\omega)=0\,{,} \quad l=1,2,\ldots \, {.}
\end{equation}
For pure imaginary frequencies  $\omega=i\omega _n$, with $\omega _n$
being the
Matsubara frequencies (\ref{Matsubara}), the left-hand sides of Eqs.\
(\ref{dd}) are given by
\begin{eqnarray}\nonumber
  \Delta_l^{\text{TE}}(ia\omega _n) &=&
  \sqrt{\varepsilon_1}s_l^{\prime}(k_{1n}a)e_l(k_{2n}a)-
  \sqrt{\varepsilon_2}s_l(k_{1n}a)e_l^{\prime}(k_{2n}a)\,, \\
  \Delta_l^{\text{TM}}(ia\omega _n) &=&
  \sqrt{\varepsilon_2}s_l^{\prime}(k_{1n}a)e_l(k_{2n}a)-
  \sqrt{\varepsilon_1}s_l(k_{1n}a)e_l^{\prime}(k_{2n}a)\,,\label{TETM}
\end{eqnarray}
where $k_{\alpha n}=\sqrt{\varepsilon_\alpha}\,\omega _n$, $\alpha =1,2$,
and $s_l(x)$,
$e_l(x)$ are the modified Riccati--Bessel functions \cite{AS}
\begin{equation}\label{RB}
  s_l(x)=\sqrt{\frac{\pi x}{2}}\,I_{\nu}(x)\,,\quad
  e_l(x)=\sqrt{\frac{2 x}{\pi}}\,K_{\nu}(x)\,,\quad
  \nu=l+\frac{1}{2}\,.
\end{equation}
The prime in Eq.\ (\ref{TETM}) stands for the differentiation with
respect to the entire argument of the Riccati--Bessel functions. The
permittivities
$\varepsilon _\alpha , \quad \alpha =1,2$ are assumed to be independent of
 the frequency
$\omega $ (dispersion is ignored) and of the temperature $T$.

Following the same way as in calculations of the Casimir energy at zero
temperature \cite{LSN} and making use of Eqs.\ (\ref{internal}),
(\ref{subtr}), and (\ref{TETM}), we obtain the internal energy
 of a dielectric ball in the form
\begin{equation}
\label{U}
U(T)= -T \sum_{l=1}^\infty (2l+1)  \sump w_n \frac{d}{d w_n}\ln \left [
W^2_l(n_1w_n,n_2w_n) - \frac{\Delta n^2}{4} P_l^2(n_1w_n,
n_2 w_n)
\right ],
\end{equation}
where
\begin{eqnarray}
  W_l(n_1w_n, n_2w_n)&=&s_l(n_1w_n)e_l^{\prime}(n_2w_n)-
s_l^{\prime}(n_1w_n)e_l(n_2w_n)\,,
  \label{W} \\
  P_l(n_1w_n, n_2w_n)&=&s_l(n_1w_n)e_l^{\prime}(n_2w_n)+
s_l^{\prime}(n_1w_n)e_l(n_2w_n)\,{,}
  \label{P}
\end{eqnarray}
 and we have introduced the dimensionless Matsubara frequencies
\begin{equation}
\label{M1}
w_n=a \omega_n = 2 \pi n a T, \quad n=0,1,2, \ldots \,{.}
\end{equation}

   It is easy to be convinced that Eq.\ (\ref{U}) can be derived from Eq.\
(2.12) in paper \cite{LSN} by the substitution
\begin{equation}
\label{sub}
dy \to 2\pi a T \sump \delta (y-w_n)\, dy\,{.}
\end{equation}

Comparing Eq.\ (\ref{free}) and Eq. (2.12) in Ref.\ \cite{LSN} one
arrives at the following inference. In order to get the free
energy at finite temperature by making use of the substitution
(\ref{sub}), one integration by parts should preliminary be done
in Eq.\ (2.12) in \cite{LSN}. Thus proceeding from the well
justified equations for the free energy (\ref{free}) and for the
internal energy (\ref{internal}) at finite temperature, one can
escape necessity to solve the problem: which energy (free or
internal) is obtained on  the substitution (\ref{sub}) in the
initial integral representation for the Casimir energy at zero
temperature \cite{BY,Kpc}.

In Eq.\ (\ref{U}) we have subtracted the contribution of an
unbounded homogeneous medium obtained in the limit $a \to \infty $.
As in the case of zero temperature, it gives, specifically, the term
linear in $\Delta n$ [see Eq.\ (2.11) in Ref. \cite{LSN}].  According to
the renormalization procedure developed in Ref.\ \cite{LSN} this contribution
should be canceled by the corresponding counter term.

     The necessity to subtract the contributions into the vacuum
energy linear in  $\varepsilon_1-\varepsilon_2$ is justified by the
following consideration.  The Casimir energy of a dilute dielectric
ball can be thought of as the net result of the van der Waals
interactions between the molecules making up the ball \cite{MNg2}.
These interactions are proportional to the dipole momenta of the
molecules, i.e., to the quantity $(\varepsilon_1-1)^2$. Thus, when a
dilute dielectric ball is placed in the vacuum, then its Casimir
energy should be proportional to $(\varepsilon_1-1)^2$. It is natural
to assume that when such a dielectric ball is surrounded by an
infinite dielectric medium with permittivity $\varepsilon_2$, then
its Casimir energy should be proportional to
$(\varepsilon_1-\varepsilon_2)^2$. The physical content of the
contribution into the vacuum energy linear in
$\varepsilon_1-\varepsilon_2$ has been investigated in the framework
of the microscopic model of the dielectric media (see Ref.\
\cite{MSS}, and references therein). It has been shown that this term
represents the self-energy of the electromagnetic field attached to
the polarizable particles or, in more detail, it is just the sum of
the individual atomic Lamb shifts. Certainly this term in the vacuum
energy should be disregarded when calculating the Casimir energy
originated in the electromagnetic interaction between
different polarizable particles or atoms
\cite{Barton,BMM,Barton-dis,H-Brevik,H-Brevik-A}.

However, there is an opposite point of view on the $\Delta n$
contribution to the vacuum energy of a pure dielectric ball according to
which this term has a real physical meaning and when calculating its value
an ultraviolet cutoff should be introduced. In this problem there is a
natural cutoff.  Really, if $d$ is a typical distance between the atoms or
molecules inside the ball then photons with energy grater than $d^{-1}$ do
not ``feel" the dielectric body and propagate freely.  This point of view
goes back to the series of papers by Schwinger who has tried to explain in
this way the sonoluminescence~\cite{Schwinger}. Further development of
this  approach can be found in Refs. \cite{Visser}. Controversy on this
subject is going on (see, for example, Ref.\ \cite{Mil-Vis}).  In any
case the $\Delta n^2$ contribution has, without doubts, real physical
meaning and it is this term that is considered below.

{}For arbitrary material media inside and outside of the ball with
permittivities $\varepsilon _1,\;\varepsilon _2$ and permeabilities
$\mu_1 \;\mu_2$, respectively, the following substitutions should be
done in Eq.\ (\ref{U})
\begin{equation}
n_i \to \frac{1}{c_i}=\sqrt{\varepsilon_i \mu_i}, \quad i=1,2\,{,}
\label{new1}
\end{equation}
\begin{equation}
\label{new2}
\frac{\Delta n^2}{4} \to \left (
\frac{\sqrt{\varepsilon_1 \mu_2}
-\sqrt{\varepsilon_2 \mu_1}}{\sqrt{\varepsilon_1 \mu_2}
+\sqrt{\varepsilon_2 \mu_1}}
\right )^2 \equiv \xi^2\,{.}
\end{equation}
With account of these substitutions it easy to do the transition to
continuous velocity of light on the surface of a compact ball placed
in material surroundings $\varepsilon
_1\mu_1=\varepsilon_2\mu_2=c^{-2}$.  In this case the internal
Casimir energy is again determined by Eq.\ (\ref{U}) with obvious
changes
\begin{eqnarray}
\label{new3}
W_l\left ( \frac{w_n}{c},\frac{w_n}{c}
\right ) &=&-1,\nonumber \\
P_l\left ( \frac{w_n}{c},\frac{w_n}{c}
\right ) &=&\left [s_l\left(\frac{w_n}{c}
\right )e_l\left (\frac{w_n}{c}
\right )
\right ]'\,{,}                \nonumber \\
\xi^2=
\left (
\frac{\sqrt{\frac{\displaystyle \ve _1}{\displaystyle \ve_2}}
-\sqrt{\frac{\displaystyle \ve _2}{\displaystyle \ve_1}}}
{\sqrt{\frac{\displaystyle \ve_1}{\displaystyle \ve_2}}
+\sqrt{\frac{\displaystyle \ve_2}{\displaystyle \ve_1}}}
\right )^2
&=&\left(\frac{\ve_1-\ve_2}{\ve_1+\ve_2}
\right)^2=\left (\frac{\mu_1-\mu_2}{\mu_1+\mu_2}
\right )^2{.}
\end{eqnarray}

Now we return to consideration of a dilute dielectric ball
and  content ourselves with the $\Delta n^2$ approximation.
In this case the contributions of $W^2_l$ and $P_l^2$ into the
internal energy (\ref{U}) are additive
\begin{equation}
\label{PW}
U(T)=U_P(T)+U_W(T)\,{.}
\end{equation}
In the approximation chosen we can put $P_l(n_1w_n, n_2w_n)\simeq
P_l(w_n,w_n)$ with the result
\begin{equation}
\label{UP}
U_P(T)=\frac{\Delta n^2}{4}T\sum_{l=1}^\infty (2l+1)\sump w_n\frac{d}{dw_n}
P_l^2(w_n,w_n)\,{.}
\end{equation}
Analysis of divergencies in the problem at hand, carried out in paper
 \cite{LSN}, leads to the
following recipe for obtaining the contribution into the internal
energy of the $W_l^2$ term in the argument of the logarithm
function in Eq.\ (\ref{U}). It is sufficient to calculate the
contribution of the $W^2_l$ term alone and then to change the sign
of this contribution to the opposite one. Hence
\begin{equation}
\label{UW}
U_W(T)= T\sum_{l=1}^\infty (2l +1) \sump w_n\frac{d}{dw_n}W_l^2(n_1w_n,
n_2w_n)\,{,}
\end{equation}
and only the term proportional to $\Delta n^2$ should be preserved in this
expression.

By making use of the addition theorem for the Bessel functions \cite{AS} the
sum over the angular momentum $l$ in Eqs.\ (\ref{UP}) and (\ref{UW}) can be
represented in a closed form in the same way as it has been done at zero
temperature in papers \cite{LSN,Klich}
\begin{eqnarray}
\sum_{l=1}^\infty (2l +1) P_l^2(w_n,w_n)&=&\frac{1}{2}\int^2_0\left[
\frac{d}{d w_n}\left (
\frac{w_n}{R}e^{-2w_n R}
\right )
\right ]^2 R\,dR-e^{-4w_n},
\label{S1} \\
\sum_{l=1}^\infty (2l+1)  W_l^2(n_1w_n,n_2w_n)&
=&\frac{\Delta n^2}{8}\int^2_{\Delta n}
\frac{e^{-2w_n R}}{R^5}(4+R^2+4w_n R-w_nR^3)^2 dR - e^{2\Delta n w_n}.
\label{S2}
\end{eqnarray}
Upon substituting  the  expressions  (\ref{S1}) and (\ref{S2}) into Eqs.\
(\ref{UP}) and (\ref{UW}), respectively, first the differentiation $d/dw_n$
should be done, and
only after that the integral over $dR$ must be evaluated. It gives
\begin{eqnarray}
 U_P(T)&=& \frac{\Delta n^2}{4}\sump \left [
\left (
2 w_n^2+2 w_n+\frac{1}{2}
\right )e^{-4w_n}  -\frac{1}{2}
\right ],
        \label{UP1} \\
U_W(T) &=&\frac{\Delta n^2}{8}T\sump\left \{(1+4 w_n)e^{-4w_n}
\right . \nonumber \\
&&- \left [  1 - 2\Delta n + \frac{16}{\Delta n}+\frac{w_n^2}{\Delta n^2}
(16-8\Delta n^2+\Delta n^4)
\right ] e^{-2 \Delta nw_n} \nonumber \\
&&\left .+ 16 w_n^2 \int^2_{\Delta n}\frac{e^{-2w_nR}}{R}dR
\right \}{.}
\label{UW1}
\end{eqnarray}
 In Eq.\
(\ref{UW1}) only the terms proportional to $\Delta n^2$ should be
preserved~\cite{LSN}, the rest of the terms being irrelevant to
our consideration.
When deriving Eq.\ (\ref{UW1}) we have  dropped the last term
in Eq.\ (\ref{S2}), $e^{2\Delta n w_n}$, which gives rise to divergence when
calculating the sum over the Matsubara frequencies.

First we consider the internal energy (\ref{UP1}). Summation over the Matsubara
frequencies can be done by making use of the formula
\begin{equation}
\label{sum}
\sump e^{-4w_n}= \frac{1}{2} \coth (4\pi aT)=\frac{1}{2}+
\frac{1}{e^{8\pi aT}-1}\,{.}
\end{equation}
It gives
\begin{equation}
\label{UPf}
             U_P(T)=
\frac{\Delta n^2}{8} T\left [
t^2\frac{\coth (2t)}{\sinh^2(2t)}+\frac{t}{\sinh^2(2t)} +\frac{1}{2}\coth (2t)
\right ], \quad
t=2\pi aT\,{.}
\end{equation}
It is worth noting that the last term  under the sum sign in Eq.\ (\ref{UP1})
gives zero contribution, when the zeta regularization technique \cite{Od,Ten}
is applied
\begin{equation}
\label{zero}
   -\frac{1}{2}\sump n^0=-\frac{1}{2}\left(
\frac{1}{2} + \zeta (0)
\right )=0\,{,}
\end{equation}
where $\zeta(z)$ is the Riemann zeta function, $\zeta(0)=-1/2$. At zero
temperature this term gives rise to a divergence that has been removed by
respective subtraction \cite{LSN}.

{}From Eq.\ (\ref{UPf}) we deduce the following behaviour of the
internal energy $U_P(T)$ at low temperature
\begin{equation}
\label{UPto0}
U_P(T)=
\frac{5\Delta n ^2}{128 \pi a}+\frac{2}{45}\Delta n^2(\pi a)^3T^4
+\frac{128}{945} \Delta n^2 (\pi a)^5 T^6- \frac{512}{525}\Delta n^2
(\pi a)^7T^8+{\cal O}(T^{10})\,{.}
\end{equation}

Integration of the thermodynamic relation (\ref{td}) enables one to get the
free energy
\begin{equation}
\label{tdint}
F(T)= - T \int \frac{U(T)}{T^2}dT + C\,T\,{,}
\end{equation}
where $C$ is a constant. Upon substitution of Eq.\ (\ref{UPf}) into
Eq.\ (\ref{tdint}) the first two terms can be integrated explicitly
\cite{KFMR}, the last term in Eq.\ (\ref{UPf}) leads to the integral
\[
  \int \frac{dx}{x}\coth (x)\,{,}
\]
which cannot be expressed in terms of the table integrals \cite{GR}. Further we
shall use  Eq.\ (\ref{tdint}) for obtaining the asymptotics of the free energy,
keeping in mind that the internal energy $U_W(T)$ in Eq.\ (\ref{UW1}) cannot be
represented in a simple closed form such as Eq.\ (\ref{UPf}).

 Substituting the asymptotics (\ref{UPto0}) into Eq.\ (\ref{tdint})
we obtain  the respective free energy
in the  low temperature region
\begin{equation}
\label{FPto0}
F_P(T)=\frac{5 \Delta n^2}{128 \pi a} - \frac{2}{135} \Delta n^2 (\pi a)^3T^4
-\frac{128}{4725}\Delta n^2 (\pi a)^2 T^6 + \frac{512}{3675} \Delta n^2
(\pi a)^7 T^8+  {\cal O}(T^{10})\,{.}
\end{equation}
Here the  linear in $T$ term $CT$ has been dropped, because the
requirement that the entropy $S_P(T)$  should vanish at $T=0$
gives~\cite{KFMR}
\begin{equation}
\label{S}
S_P(0)=\lim_{T\to 0} T^{-1} \left (
U_P(T)-F_P(T)
\right )=C=0\,{.}
\end{equation}
Hence, at low temperature the expansions both for the internal energy
(\ref{UPto0}) and for the free energy (\ref{FPto0}) involve only even powers of
the temperature beginning from $T^4$.
At zero temperature we have
\begin{equation}
\label{T0}
U_P(0)=F_P(0)=E_P=\frac{5\Delta n^2}{128 \pi a}\,{,}
\end{equation}
where $E_P $ is the Casimir energy of a compact ball with the same
velocity of light inside and outside the ball \cite{LSN,Klich,BD}.

Our calculation of the free energy $F_P(T)$ corresponds to the two-scattering
approximation in treatment of a perfectly conducting spherical shell in Ref.\
\cite{BD}. The relevant results of that paper should be multiplied by
$\Delta n^2/4$ before comparing with ours. However, the free energy of a
conducting sphere, calculated
in the two-scattering approximation, is presented there only graphically, and
Eq.\ (8.37) in that paper gives the low temperature behavior of an exact
result for this quantity. Therefore the coefficients in this equation are
a bit different
as compared with the  two first terms in our Eq.\ (\ref{FPto0}).

When temperature $T$ tends to infinity, Eq.\  (\ref{UPf}) gives
\begin{equation}
\label{UPtoinfty}
U_P(T)\simeq \frac{\Delta n^2}{16}\,T, \quad T \to \infty\,{.}
\end{equation}

Substituting this asymptotics into Eq.\ (\ref{tdint}) we arrive at the high
temperature limit for the free energy $F_P(T)$
\begin{equation}
\label{FPtoinfty}
F_P(T)\simeq - \frac{\Delta n^2}{16} T \left [\ln (aT) + \alpha \right ],
\quad T \to \infty
.
\end{equation}
 The constant $\alpha $ can be find by making use of the complete expression
for $F_P(T)$ (see Refs.\ \cite{KFMR,NPB})
\[
\alpha =\gamma +\ln 4 - \frac{5}{4}\,{.}
\]

We have explained above how to do the transition to continuous
velocity of light on the surface of a material ball [see Eqs.\
(\ref{new1}--(\ref{new3})]. With allowance for this we immediately
conclude that the internal energy $U_P(T)$ and free energy $F_P(T)$
exactly concern that configuration, certainly upon the substitution
(\ref{new2}). Our functions $U_P(T)$ and $F_P(T)$ completely coincide
with calculations in Refs.\ \cite{Klich,KFMR} where the addition
theorem for the Bessel functions has been applied also. But in our
problem (a pure dielectric ball) there is an additional contribution
to the vacuum energy generated by the functions $W_l^2$ in Eq.\
(\ref{U}). Now we turn to the analysis of this contribution.


The summation over
the Matsubara frequencies in Eq.\ (\ref{UW1}) gives
\begin{eqnarray}\label{UWf}
  U_W(T)& = & \frac{\Delta n^2}{8}\, T\,
  \left\{\frac{1}{e^{4t}-1}\left(1+
\frac{4t}{1-e^{-4t}}\right)-\frac{1}{e^{2\Delta n
  t}-1}\right. \nonumber \\
     & &  -\frac{t}{2\sinh^2 (\Delta n t)}\left[\frac{8}{\Delta
    n}-\Delta n -2t\left(2-\frac{4}{\Delta n^2}-\frac{\Delta
    n^2}{4}\right)\coth (\Delta n t)\right]  \nonumber \\
     & & \left. +t^2\int_{\Delta n}^2 \frac{dR}{R}\frac{\coth (tR)}{\sinh^2
    (tR)}\right\}\,,\quad t=2\pi aT {.}
\end{eqnarray}

According to the renormalization procedure employed, only the
terms proportional to $\Delta n^2$ should be retained in Eq.\
(\ref{UWf}). Obviously, this can be accomplished explicitly
when considering the asymptotics of the internal energy $U_W(T)$
for low and high temperatures. For low $T$ Eq.\ (\ref{UWf}) gives
\begin{eqnarray}\label{UWto0}
  U_W(T) & = & \frac{\Delta n^2}{48 \pi a}-\frac{16}{45} \, \Delta
  n^2 (\pi a)^2\,T^4+\frac{1024}{2835}\,\Delta n^2 (\pi a)^5\, T^6
\nonumber     \\
  & & -\frac{4096}{7875}\, \Delta n^2 (\pi a)^7\, T^8+{\cal O}(T^{10})\,.
\end{eqnarray}
By making use of Eq. (\ref{tdint}) with the constant $C$ equal to
zero, we obtain the low temperature expansion for the respective
free energy
\begin{eqnarray}\label{FWto0}
  F_W(T) & = & \frac{\Delta n^2}{48 \pi a}+\frac{16}{135}\,
  \Delta n^2 (\pi a)^3\, T^4-\frac{1024}{14175}\,\Delta n^2 (\pi a)^5\, T^6
\nonumber \\
  & & + \frac{4096}{55125}\, \Delta n^2 (\pi a)^7\, T^8+{\cal O}(T^{10})\,.
\end{eqnarray}
The sum of Eqs. (\ref{UPto0}) and (\ref{UWto0}) gives the total
internal energy at low temperature
\begin{eqnarray}\label{Uto0}
 U(T) & = & \frac{23}{384}\, \frac{\Delta n^2}{\pi
 a}-\frac{14}{45}\, \Delta n^2 (\pi a)^3\, T^4+\frac{1408}{2835}\,
 \Delta n^2 (\pi a)^5 \, T^6
 \nonumber \\
 & & -\frac{11776}{7875}\, \Delta n^2(\pi a)^7\, T^8+{\cal O}(T^{10})\,.
\end{eqnarray}
{}From Eqs. (\ref{FPto0}) and (\ref{FWto0}) we obtain for the total free
energy of a
dilute dielectric ball
\begin{eqnarray}\label{Fto0}
  F(T) & = & \frac{23}{384}\frac{\Delta n^2}{\pi
  a}+\frac{14}{135}\, \Delta n^2 (\pi a )^3\, T^4  \nonumber \\
  & & -\frac{1408}{14175}\,\Delta n^2 (\pi a)^5\,
  T^6+\frac{11776}{55125}\,\Delta n^2 (\pi a)^7\,
  T^8+{\cal O}(T^{10})\,.
\end{eqnarray}

The first three terms of the asymptotics~(\ref{Fto0}) has been
derived in a recent paper by Barton~\cite{Barton-T} in the
framework of a completely different approach, namely by making use of
perturbative theory for quantized electromagnetic field where
dielectric ball is considered as a perturbation in unbounded
continuous surroundings. When comparing our Eq.\ (\ref{Fto0}) with
respective formula in the Barton paper \cite{Barton-T} one should
take into account that our quantity $\Delta n^2$ is equal to  the
Barton's $4 \pi^2 (n \alpha)^2$. In Ref.\ \cite{Barton-T} an
additional term proportional to $T^3$ has been obtained for the free
energy in the low temperature limit. It should be noted that the
$T^3$ term does not give contribution to the Casimir pressure exerted
on the surface of a dielectric ball. In this respect this term is
nonobservable. In our approach the $T^3$ term is absent because at
first we have calculated the total energy $U(T)$ and after that we
derive the free energy $F(T)$ proceeding from $U(T)$.
At zero temperature we have
\begin{equation}\label{TT0}
  U(0)=F(0)=E=\frac{23}{384}\frac{\Delta n^2}{\pi a}\,,
\end{equation}
where $E$ is the Casimir energy of a dilute dielectric ball
calculated in Ref.\ \cite{LSN}.

When passing from a compact ball with uniform velocity of light to
a pure dielectric ball, the sign of the first temperature
correction ($\sim T^4$) to the free energy and consequently to the
Casimir forces changes to the opposite one (see Eqs. (\ref{FPto0})
and (\ref{Fto0})). Keeping two terms in the expansion
(\ref{Fto0}) we get for the Casimir forces exerted on the unit
area of the ball
surface
\begin{equation}
\label{force} {\cal F} = -\frac{1}{4 \pi a^2}\frac{\partial F(T)}{\partial
a}=\frac{23}{1636} \frac{\Delta n^2}{\pi^2a^4}\left [ 1-
\frac{112}{345}(2\pi a T)^4
\right ]{.}
\end{equation}

{}From Eq.\ (\ref{UW1}) one can derive the high temperature behaviour of the
internal energy $U_W(T)$ in the same way as Eq.\ (\ref{UPtoinfty}) has
been obtained
\begin{equation}
\label{UWtoinfty}
U_W(T)\simeq \frac{\Delta n^2}{8}T\frac{1}{2}=\frac{\Delta n^2}{16} T,
\quad T  \to \infty {.}
\end{equation}
Substitution of this result into Eq.\ (\ref{tdint}) and subsequent integration
gives
\begin{equation}
\label{FWtoinfty}
F_W\simeq -\frac{\Delta n^2}{16}T\left [\ln (aT)+\beta \right ]\,{,}
\end{equation}
where $\beta $ is a constant.

By making use Eqs. (\ref{UPtoinfty}), (\ref{FPtoinfty}), (\ref{UWtoinfty}),
and (\ref{FWtoinfty}) we obtain the high temperatute asymptotics of the
internal energy and free energy of a dilute dielectric ball
\begin{eqnarray}
 U(T) &=& U_P(T)+U_W(T) \simeq \frac{\Delta n^2}{8}\, T\,,
\quad T\to \infty {,}  \label{Utoinfty}
 \\
 F(T)&=&F_P+F_W(T)  \simeq   -\frac{\Delta n^2}{8}\, T
\left [\ln (aT)+c\right ] \,, \quad T\to \infty\,{,}
\label{Ftoinfty}
\end{eqnarray}
where $c=\alpha +\beta $ is a constant the exact value of wich can be obtained
by other methods~\cite{Barton-T,NPB}
\begin{equation}
\label{c}
c=\ln 4+\gamma-\frac{7}{8},
\quad \beta =\frac{3}{8}{.}
\end{equation}
Exactly the same high temperature behaviour of the thermodynamic functions
of electromagnetic field connected with a dilute dielectric ball
has been obtained in Ref.\ \cite{Barton-T}.

 With allowance for the dimension of the respective quantities it is
easy to be convinced that Eq.\ (\ref{Utoinfty}) and the last term in Eq.\
(\ref{Ftoinfty})
do not contain the Planck constant. Hence, the high temperature limit
for the internal and free energies implies the classical limit
\cite{Milton,FMR}. At the same time these leading terms
do not depend on the radius of the ball $a$ too and, as a result, they do
not contribute to  the Casimir force at high temperature.
From the asymptotics (\ref{Uto0}),  (\ref{Fto0}),
(\ref{Utoinfty}), and (\ref{Ftoinfty}) it follows that the
thermodynamic characteristics $U(T)$ and $F(T)$ of a dilute
dielectric ball have, respectively, minimum and maximum at
nonzero values of the temperature~$T$.

The characteristic temperature scale for the thermodynamic quantities under
consideration is determined by the radius $a$ of the ball. For $a
 \sim 10^{-4}$~cm this
scale proves to be large $\sim {1000\,}^\circ $K.

\section{Conclusion}
In this paper the Casimir internal and free energies are
calculated  for a dilute dielectric ball at finite temperature. As
we are aware, it has been done for the first time. The explicit
formulas are derived which allow one to develop the expansions for
thermodynamic characteristics of the ball at low and high
temperatures. It is found that the first temperature correction
 $(\sim T^4)$ to the free energy in the problem at hand has an opposite sign
as compared with a perfectly conducting sphere \cite{BD} and a compact ball
with constant velocity of light inside the ball and in the
surroundings\cite{BY,KFMR}.
 It implies that the Casimir force, exerted on the
surface of a dielectric ball and tending to expand it, diminishes
with rising temperature [see Eq.\ (\ref{force})].
 Usually the temperature dependence of the
Casimir forces is opposite \cite{MT,Milton,FMR}. However, for a
perfectly conducting
wedge in the low temperature region the Casimir forces also decrease when the
temperature rises \cite{BD}.

Without doubt,  it is worth considering this problem in the framework
of other methods, for example, by making use of the Green's
function techniques. However, before doing this the role of the
so-called contact terms in the expression of the vacuum energy
employed there should be elucidated. In our global approach these
terms do not appear because we are only  dealing  with the sum of
the eigenfrequencies of the electromagnetic field under given
boundary conditions.

\acknowledgments
V.V.N.~thanks Professor Barton for providing his  paper \cite{Barton-T}
and for very fruitful communications.
This research has been supported by the fund MURST ex 40\% and 60\%,
art.\ 65 D.P.R.\  382/80.  The work was  accomplished during the visit
of V.V.N. to Salerno University. It is a pleasure for him to
thank Professor G.\ Scarpetta, Drs.  G.\ Lambiase and A. \ Feoli for
warm hospitality. The financial support of IIASS is acknowledged.
G.L.\ thanks the UE, P.O.M.\ 1994/1999, for financial
support.  V.V.N.\ acknowledges the partial financial support of
Russian Foundation for Basic Research (Grant No.\ 00-01-00300).
G.L.\ and V.V.N. are indebted to I.\ Klich for useful discussions.

\end{document}